\documentclass[preprint2]{aastex}

\newcommand{\jh}{\textit{J--H}}
\newcommand{\hk}{\textit{H--K}}
\newcommand{\kl}{\textit{K--L}}
\newcommand{\ji}{\textit{J}}
\newcommand{\hi}{\textit{H}}
\newcommand{\ki}{\textit{K}}
\newcommand{\li}{\textit{L}}
\newcommand{\ksi}{\textit{K}\ensuremath{_{s}}}
\newcommand{\lpi}{\textit{L}\ensuremath{^{\prime}}}
\newcommand{\kc}{\textit{K-continuum}}
\newcommand{\htwo}{H\ensuremath{_{2}}~\ensuremath{v=1-0}~S(1)}
\newcommand{\htw}{H\ensuremath{_{2}}}
\newcommand{\kt}{\ensuremath{k_{B}T}}
\newcommand{\lx}{\ensuremath{L_{X}}}
\newcommand{\nh}{\ensuremath{N_{\rm H}}}

\shorttitle{Deep NIR Observations of the X-ray Emitting Class 0 Protostar Candidates in the OMC-3}
\shortauthors{Tsujimoto et al.}

\begin{document}
\title{Deep Near Infrared Observations of the X-ray Emitting Class 0 Protostar Candidates in the Orion Molecular Cloud-3\altaffilmark{1}}
\author{Masahiro Tsujimoto, and Katsuji Koyama}
\affil{Department of Physics, Graduate School of Science, Kyoto University, Sakyo-ku, Kyoto, 606-8502, Japan}
\email{tsujimot@cr.scphys.kyoto-u.ac.jp, koyama@cr.scphys.kyoto-u.ac.jp}
\author{Yohko Tsuboi, and George Chartas}
\affil{Department of Physics \& Astrophysics, 525 Davey Laboratory, Pennsylvania State University, University Park, PA 16802, USA}
\email{tsuboi@astro.psu.edu, chartas@astro.psu.edu}
\author{Miwa Goto\altaffilmark{2,3}, Naoto Kobayashi\altaffilmark{2}, and Hiroshi Terada}
\affil{Subaru Telescope, National Astronomical Observatory of Japan, 650 North A'ohoku Place, Hilo, HI 96720, USA}
\email{mgoto@subaru.naoj.org, naoto@subaru.naoj.org, terada@subaru.naoj.org}
\and
\author{A. T. Tokunaga}
\affil{Institute for Astronomy, University of Hawaii, 2680 Woodlawn Drive, Honolulu, HI 96822, USA} 
\email{tokunaga@ifa.hawaii.edu}
\altaffiltext{1}{Based on data collected at Subaru Telescope, which is operated by the
National Astronomical Observatory of Japan.}
\altaffiltext{2}{Visiting Astronomer at the Infrared Telescope Facility, which is
operated by the University of Hawaii under contract from the National Aeronautics and
Space Administration.}
\altaffiltext{3}{Visiting Astronomer at the Institute for Astronomy, University of Hawaii}

\begin{abstract}
We obtained near infrared (NIR) imaging with the \textit{Subaru} Telescope of the
 class~0 protostar candidates in the Orion Molecular Cloud-3, two of which were
 discovered to have X-ray emission by the \textit{Chandra} X-ray Observatory. We found
 strong evidence for the class~0 nature of the X-ray sources. First, our deep \ki-band
 image shows no emission brighter than 19.6~mag from both of these X-ray sources. Since
 class~I protostars or class~II T Tauri stars should be easily detected in the NIR with
 this sensitivity, the lack of \ki-band detection suggests that they are likely much
 more obscured than class~I protostars. Second, our \htwo\ image shows a bubble-like
 feature from one of the X-ray class~0 protostar candidates, which reinforces the idea
 that this is a class~0 protostar. We also discuss the nature of nine NIR sources found
 in our deep image based on their colors, spatial coincidence with millimeter cores, and
 the properties of their X-ray counterparts.
\end{abstract}
\keywords{stars: pre-main sequence --- Infrared: stars --- X-rays: stars --- Individual: OMC-3}

\section{INTRODUCTION}
Extensive mapping observations at various wavelengths have been performed in the Orion
Molecular Cloud 2 and 3 (hereafter OMC-2/3) in the search for protostars. \citet{chini97}
detected a chain of 21 cores (MMS1--MMS10 in OMC3, FIR1a--FIR6d in OMC-2) at 1.3~mm
using Institut de Radio Astronomie Millim\'{e}trique (IRAM) telescope. Combining the
James Clerk Maxwell Telescope (JCMT) photometry from 350 \micron\ to 2~mm and the
Infrared Astronomical Satellite (IRAS) photometry from 12 \micron\ to 100 \micron, 
they showed that the sub-millimeter luminosity (L$_{\rm{submm}}$) of most cores
dominates the bolometric luminosity (L$_{\rm{bol}}$). \citet{reipurth99} observed the
same region at 3.6~cm using Very Large Array (VLA), and reported the detection of eight
1.3~mm cores. \citet{lis98} discovered more than 30 cores (CSO1--CSO33) in OMC-2/3 at 350
\micron, and found that all but one of the 1.3~mm cores have counterparts in 350
\micron. \citet{johnstone99} obtained images of this region in 450~\micron\ and
850~\micron. Further, \citet{yu97} studied this region with \htwo\ (hereafter \htw)
emission line and found that about half of the 1.3~mm cores are associated with \htw\
jets. \citet{aso00} found that some cores are also accompanied by CO and HCO$^{+}$
molecular outflows. All these indicate the existence of protostars, most of which are
probably at the class 0 stage.

More astonishing is the detection of hard X-ray emission from two (MMS2 and MMS3) of the
millimeter cores using the \textit{Chandra} X-ray Observatory \citep{tsuboi01}. Due to
the spatial coincidence with millimeter cores and their hard X-ray spectra,
\cite{tsuboi01} proposed that they are the first candidates of the X-ray emitting class~0
protostars. 

Previous radio and near infrared (NIR) observations indicate the existence of class~0
protostars at MMS2 and MMS3. However, their spatial resolution was not high enough to
determine which X-ray source coincides with the class~0 protostar, particularly at MMS2
where there is a cluster of four X-ray sources. Clearly, follow-up observations with
much higher spatial resolution, comparable to that of \textit{Chandra} (positional
accuracy of $\sim$0\farcs1), are needed.

This paper reports the results of high resolution NIR imaging targeting the X-ray
sources seen at MMS2 and MMS3.

\section{OBSERVATIONS}
Observations were performed using the Infrared Camera and Spectrograph (IRCS) at the
Cassegrain focus of the \textit{Subaru} Telescope \citep{tokunaga98,kobayashi00}. We took
three broad-band (\ji, \hi, and \ki-band) and two narrow-band (\htw\ and \kc) images on
2000 November 30 and December 4. The seeing was $\sim$ 0\farcs5 on both nights. The
\ji, \ki, \htw, and \kc\ exposure times were 600 seconds each, while the \hi\ exposure time
was 300 seconds. IRCS provides a field of view (FOV) of 60\arcsec $\times$ 60\arcsec\
with a pixel scale of 0\farcs058. With dithering, we covered a 90\arcsec $\times$
90\arcsec\ field encompassing both MMS2 and MMS3.

\placefigure{fg:f1}

As we had no detection in \ji-band from two NIR sources at MMS2, we obtained an
additional \lpi-band image of MMS2 with NSFCam \citep{leggett96} at the Cassegrain focus
of the Infrared Telescope Facility (IRTF) on 2000 December 23. The seeing was $\sim$
1\farcs0 and the integration time was 216 seconds. NSFCam provides a 38\arcsec
$\times$38\arcsec\ FOV with the pixel scale of 0\farcs148. With dithering, we obtained a
64\arcsec $\times$ 64\arcsec\ field.

The images were reduced following the standard procedures using \texttt{IRAF}\footnote{IRAF
is distributed by the National Optical Astronomy Observatories, which are operated by
the Association of Universities for Research in Astronomy, Inc., under cooperative
agreement with the National Science Foundation.} : dark-subtraction, flat-fielding,
sky-subtraction, and bad pixel removal for each frame, and correction for dithering to
construct a final image.

\section{RESULTS}
\subsection{Source Extraction and Photometry of Broad-band Images}
\texttt{SExtractor} \citep{bertin96} was used for source extraction and photometry. Nine
sources (IRS1--IRS9) were extracted from the \ki-band image. (Table \ref{tb:t1}). For
each \ki-band detected source, we performed a 1\farcs0-aperture photometry of \ji, \hi,
and \ki-bands. We transformed their magnitudes into the CIT system in the following
way. Seven sources are listed in the Point Source Catalog of the 2MASS Second
Incremental Data Release\footnote{http://www.ipac.caltech.edu/2mass/}. Referring to
their \ji, \hi, and \ksi\ magnitudes, we derived a linear relation between IRCS and
2MASS magnitudes in each band. We first converted the IRCS magnitudes into 2MASS
magnitudes using these relations and then into the CIT system using the formule given
in \cite{carpenter01}.

For the \lpi-band image with NSFCam, we performed a 2\farcs0-aperture photometry of IRS3,
IRS4, and IRS5. We first calculated the magnitudes with the photometric zeropoint of
20.3~mag\footnote{http://irtf.ifa.hawaii.edu/Facility/nsfcam/hist/backgrounds.html}, and
then converted them into the CIT \li-band color using
\begin{equation}
 (\ki-\li)_{\rm{CIT}} = 0.820 \times (\ki-\lpi)_{\rm{IRTF}}\footnote{http://irtf.ifa.hawaii.edu/Facility/nsfcam/hist/color.html}
\end{equation}
where we assume \ki$_{\rm{CIT}}=$\ki$_{\rm{IRTF}}$ as the first order approximation.

\subsection{Correlation with X-ray Sources}
The X-ray counterpart was searched for each NIR source using the \textit{Chandra} data
given in \cite{tsuboi01}. From the visual inspection of the NIR
and X-ray images, the X-ray sources 6, 7, 2, 12, and 3 in \citet{tsuboi01} (hereafter we
call them TKH~6, TKH~7, TKH~2, TKH~12, and TKH~3 ; the rests follow the same rule.) were
identified to be the counterpart of IRS1, IRS4, IRS6, IRS8, and IRS9, respectively.

Two NIR sources (IRS3 and IRS5) and four X-ray sources (TKH~8, TKH~8a, TKH~8b, and TKH~8c) are
found at MMS2. In order to find the X-ray counterpart of the NIR sources, we adjusted
the X-ray image by shift and rotation so that each X-ray source (TKH~2, TKH~3, TKH~6,
TKH~7, and TKH~12) comes closest to its NIR counterpart. After this procedure, the
positional offset between the NIR sources and their X-ray counterparts is $\sim$
0\farcs25 rms (1$\sigma$).

Then, TKH~8c is found to be the closest source to IRS5 with separation of 0\farcs46,
hence is the X-ray counterpart of IRS5. On the other hand, IRS3 is separated by
0\farcs81 from the closest X-ray source; TKH~8. Assuming that the separation between a NIR
and X-ray counterpart pair follows a gaussian distribution of $\sigma=$ 0\farcs25, the
separation between IRS3 and TKH~8 is more than 3$\sigma$. We therefore conclude that
TKH~8 is not the X-ray emission from IRS3. In fact, no separation larger than 0\farcs81
is found in any other NIR and X-ray counterpart pairs. TKH~10 at MMS3, as well as TKH~8
at MMS2, has no NIR counterpart.

We chose several source-free regions near the positions of TKH~8 and TKH~10 for a
1\farcs0-aperture photometry in order to estimate the 1$\sigma$ level of the
background. We find the \ki-band upper limit of TKH~8 and TKH~10 to be $\sim$ 19.6~mag
at the 3$\sigma$ level.

\subsection{Narrow-band Images}
The vibrational-rotational transition of $v=1-0~S(1)$ works as an effective coolant of the
excited hydrogen molecules. Therefore this emission line serves as a powerful tool to
search for jets from a protostar and the position of its powering source
\citep{bally93,hodapp95}. In a continuum-subtracted \htw-band image, we identified a
bubble-like feature originating from MMS2. A close-up view of this bubble-like emission
is shown in Figure \ref{fg:f2}, where we see the origin of this feature spatially
coincides with TKH~8. No similar feature was found for TKH~10 at MMS3.

\placefigure{fg:f2}

\section{DISCUSSION}
\subsection{The Nature of NIR Sources}
For the classification of Young Stellar Objects (YSOs), we use the NIR color-color
diagram \citep{lada92}. The \jh/\hk\ diagram is given in Figure \ref{fg:f3}~(a). 
Since IRS3 and IRS5 have no positive detection in the \ji-band, we also give \hk/\kl\
diagram in Figure \ref{fg:f3}~(b). 

In these diagrams, protostars and classical T Tauri stars (CTTS) are positioned right of
the reddening vector where reddened photospheres by extinction are inaccessible,
because they have redder colors due to an intense circumstellar disk emission. Thus they
can be discriminated from weak-line T Tauri stars (WTTS). Protostars and CTTS can be
separated from each other based on the amount of extinction. Protostars generally show
larger extinction due to their richer circumstellar material than CTTS. 

\placefigure{fg:f3}

\begin{figure}
 \figurenum{1}
 \epsscale{0.65}
 \plotone{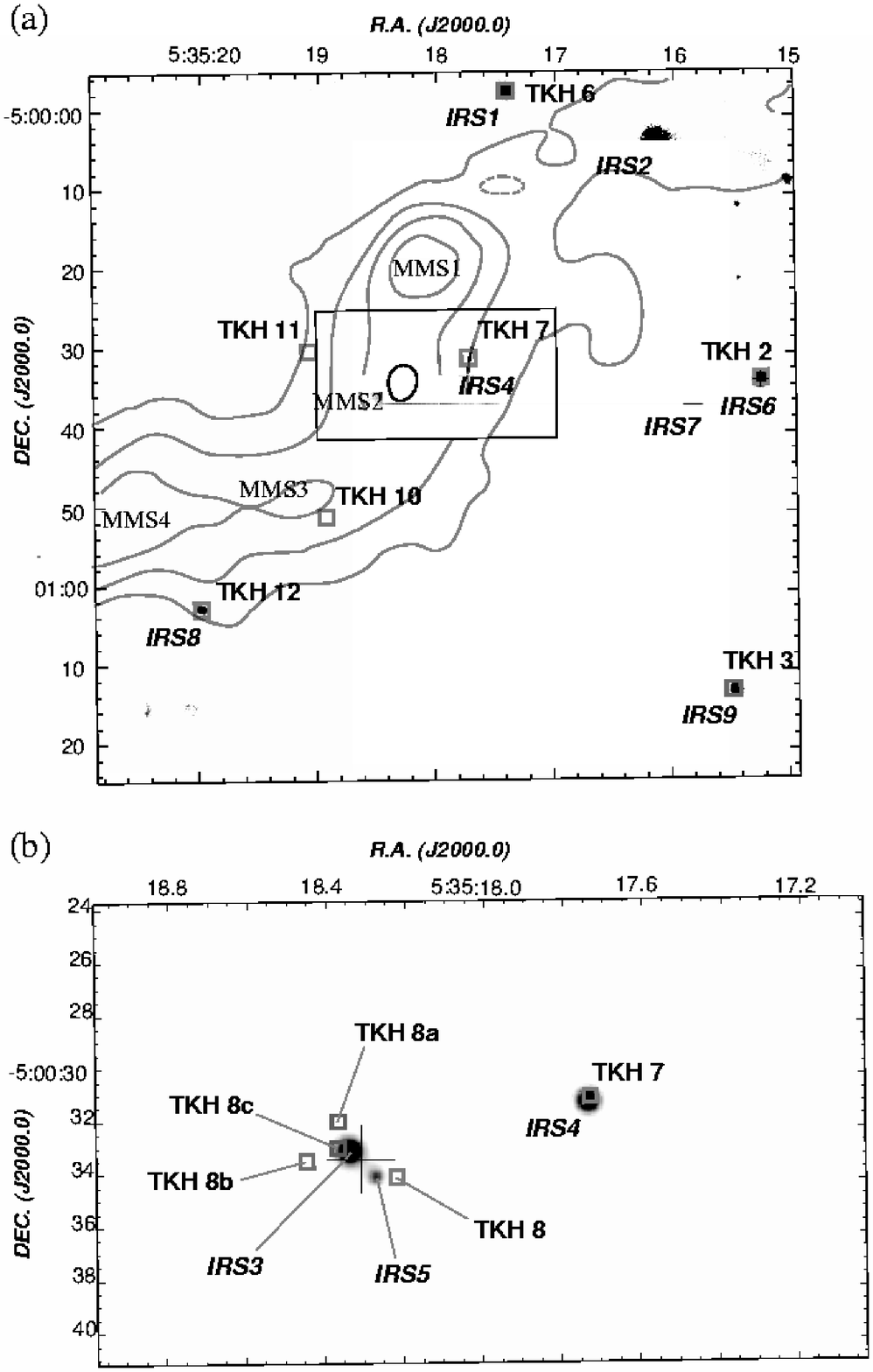}
 \caption{(a) The IRCS \ki-band image with the logarithmic gray scale to
 stress diffuse features. (b) The close-up view of the MMS2 region (shown in a rectangle
 in (a)) in the linear scale to show the accurate positions of point sources. The
 \ki-band sources (IRS1--IRS9) are labeled in italic, while the positions of the X-ray
 sources \citep{tsuboi01} are in squares with their names in roman. Contours in (a) are
 1.3~mm intensity. Four 1.3~mm cores (MMS1--MMS4) are identified in this region
 \citep{chini97}. A cross in (b) shows the position of the 3.6~cm source.\label{fg:f1}}
\end{figure}

\begin{figure}
 \figurenum{2}
 \epsscale{1.0}
 \plotone{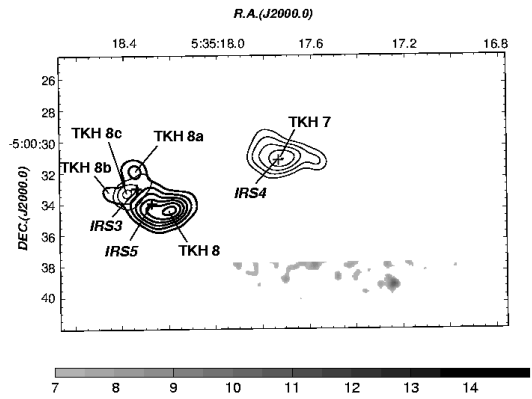}
 \caption{The continuum-subtracted \htw\ intensity (gray scale) with hard (3.0--6.0~keV)
 and soft (0.5--3.0~keV) X-ray intensity overlaid with thick and thin contours. The \kc\
 image is multiplied by a factor and subtracted from \htw\ image, so that the emissions
 from IRS3 and IRS5 cancel out. Without \kc\ subtraction, however, we confirmed the same
 bubble-like feature in the \htw\ image. The scale bar at the bottom is in the unit of
 intensity pixel$^{-1}$, where the background level (white) is $\sim$2.5. The
 position of \ki-band sources are shown with crosses. X-ray sources and NIR sources are
 labeled in roman and italic respectively.\label{fg:f2}}
\end{figure}

\begin{figure}
 \figurenum{3}
 \epsscale{0.6}
 \plotone{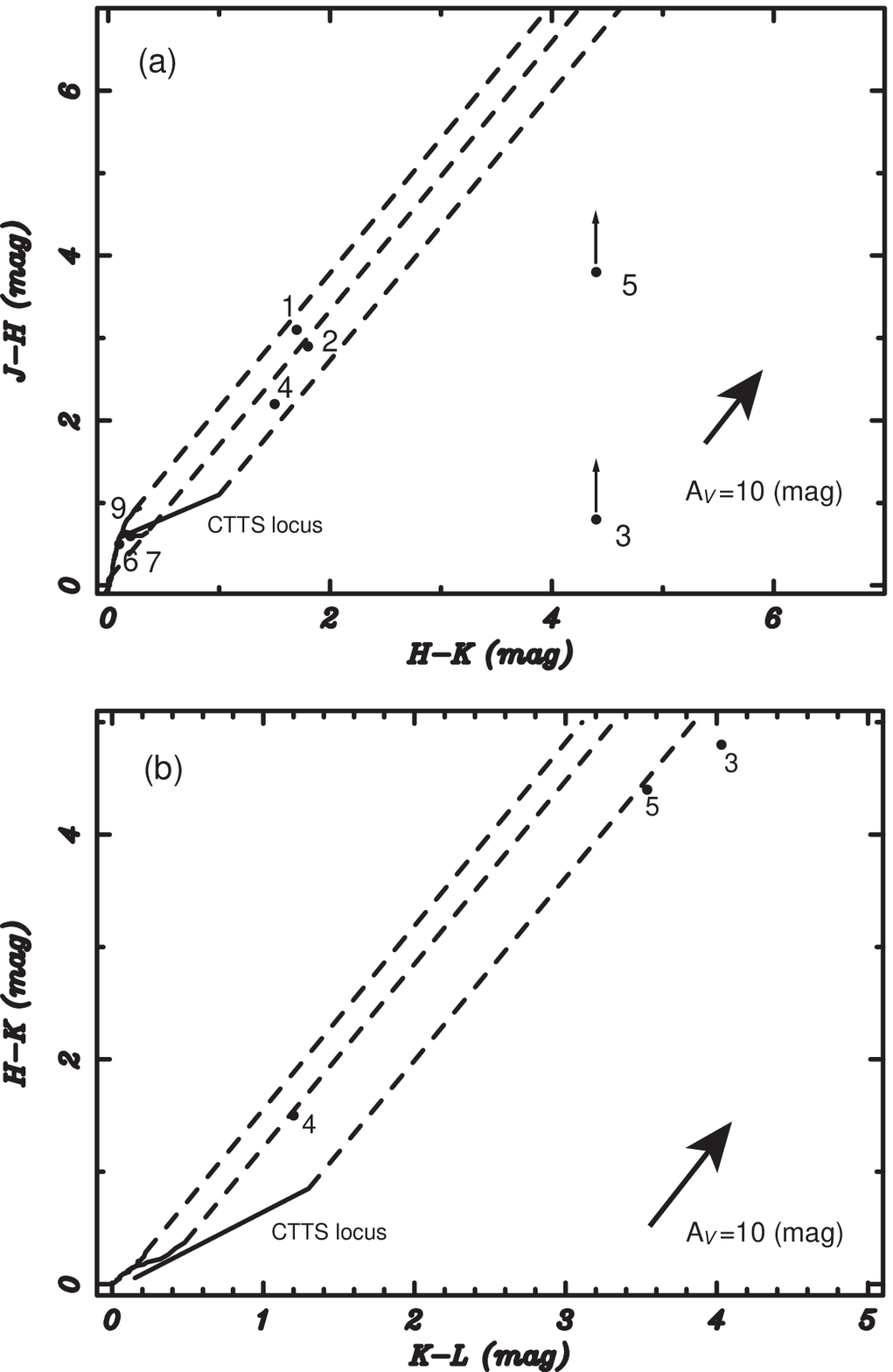}
 \caption{(a) \jh/\hk\ color-color diagram and (b) \hk/\kl\ color-color
 diagram. IRS1--IRS9 are plotted in the CIT system with the label of their names
 (``IRS'' is omitted). Errors are $\lesssim \pm$ 0.1~mag for each color. The intrinsic
 colors of dwarfs and giants \citep{tokunaga00}, and the CTTS locus \citep{meyer97} are
 shown with the solid lines, while their extinction vectors are in the dashed lines. We
 assume the slope of the reddening lines to be
 $E(\jh)_{\rm{reddening}}/E(\hk)_{\rm{reddening}}=1.69$ and
 $E(\hk)_{\rm{reddening}}/E(\kl)_{\rm{reddening}}=1.63$ \citep{meyer97}. The $A_{V}$ of
 each source is estimated from $E(\hk)_{\rm{reddening}}=0.065 \times A_{V}$ or
 $E(\kl)_{\rm{reddening}} = 0.04 \times A_{V}$ \citep{meyer97}.\label{fg:f3}}
\end{figure}

IRS3 and IRS5 are at the very center of a millimeter core (MMS2) and are located
$\sim$1\farcs34 ($\sim$ 600~AU at the distance of 450~pc) apart from each other (Figure
\ref{fg:f1}). Together with their large extinction of more than $A_{V} > 50$ mag and
large NIR excess seen in Figure \ref{fg:f3} (b), they are class I protostars probably
comprising a binary system.

IRS1, IRS2 and IRS4 are at the reddening region of the CTTS locus with a moderate
extinction of $A_{V} \sim 30$ mag (Figure \ref{fg:f3}a). They are located at the edge
of 1.3~mm cores (Figure \ref{fg:f1}), thus are most likely to be CTTS.

IRS6, IRS7, and IRS9 are located away from the cloud cores (Figure \ref{fg:f1}) and have
less extinction (Figure \ref{fg:f3}a). Among them, IRS6 and IRS9 are considered to be
WTTS due to the association with X-ray emission. IRS7, which has no X-ray counterpart,
may be a background source.

It is hard to infer the nature of IRS8 from NIR observation alone because it has only
\ki-band detection. However, its X-ray counterpart (TKH~12) shows a thermal emission of
\kt$=$3.2~keV, \lx$=5 \times 10^{30}$erg~sec$^{-1}$, and \nh$=$6$\times
10^{22}$cm$^{-2}$ \citep{tsuboi01}. According to \cite{tsujimoto01} who derived a
typical X-ray spectrum for protostars (class~I), CTTS (class~II), and WTTS (class~III)
in OMC-2/3, TKH~12 ($=$IRS8) is most likely to be a class~I protostar.

\subsection{The X-ray Emitting Class~0 Protostar Candidates}
Class 0 protostars are empirically characterized with the following features
\citep{barsony94} : 1)~undetectable at wavelengths $<$10 \micron, 2)~a high
ratio of L$_{\rm{submm}}$/L$_{\rm{bol}}$, 3)~a relatively narrow spectral energy
distribution resembling that of a single blackbody at T$\leq$30~K, 4)~presence of a
molecular outflow, and additionally 5)~existence of cm-wave continuum emission, and
6)~presence of Herbig-Haro (HH) objects.

MMS2 and MMS3 were already found to have a high ratio of L$_{\rm{submm}}$/L$_{\rm{bol}}$
\citep{chini97}. \citet{tsuboi01} proposed that hard X-ray sources at MMS2 and MMS3 are
from class~0 protostars. 

With a deep \ki-band imaging observation, we found that TKH~8 at MMS2 and TKH~10 at MMS3
have no NIR emission brighter than $\sim$ 19.6 mag. Class~I protostars are generally
detected at 10--14~mag and class~II T Tauri stars are at 8--12~mag in \ki-band at the
distance of 450~pc (for example, see Fig 24 in Aspin, Sandell, \& Russel 1994). Class~I
and class~II stars are therefore easily detected in the NIR with our sensitivity. In
fact, two class~I protostars at MMS2 (IRS3 and IRS5) show 13.2~mag and 11.4~mag in the
\ki-band. TKH~8 and TKH~10 are fainter than these stars by more than 100 times. This
suggests that they are much more obscured than class~I protostars, namely at class~0.

A bubble-like emission in \htw\ from TKH~8 reinforces this idea. \cite{yu97} identified
the \htw\ emission from MMS2 in a global scale. MMS2 is also accompanied by HH objects
\citep{reipurth97}, molecular outflows \citep{aso00}, and 3.6~cm continuum
emission \citep{reipurth99}. Our bubble-like feature is at the origin of these jet and
outflow phenomena. Therefore, this \htw\ emission should be shock-excited by the jet
from a protostar. The close-up view of this feature shows that the bubble has a length
of $\sim$ 10\arcsec\ ($= 4.5 \times 10^{3}$~AU) westward away from its central
source. This morphology is observed in jets from class~0 protostars ( e.g.,
L~1448-mm and L~1448 IRS~2 ) \citep{davis94,eisloeffel00} . The well-collimated jet is
typical of class~0's, of which the surrounding material is still rich and the
outflow can only go through narrow cavities near the pole \citep{hodapp98}. TKH~8, the
powering source of the jet, is thus most likely to be the class~0 protostar.

\section{SUMMARY}
\begin{enumerate}
 \item In deep NIR imaging of the OMC-3 region, we detected nine \ki-band sources, and
       derived their \ji, \hi, and \ki-band magnitude. We also correlated them with the
       X-ray sources and found that six sources have the X-ray counterpart.
 \item Based on their \jh, \hk, and \kl\ colors, spatial coincidence with millimeter
	cores, and X-ray properties of their X-ray counterpart, we find that three are
       class~I protostars, three are CTTS, two are WTTS, and one is a background source.
 \item With a deep \ki-band image, TKH~8 and TKH~10 were found to have no \ki-band emission
       brighter than 19.6~mag, suggesting that they are highly obscured sources than
       class~I protostars, namely class~0 protostars. With a \htwo\ image, a bubble-like
       feature was found to originate from TKH~8, reinforcing the class~0 nature of this
       source.
\end{enumerate}

\acknowledgments
We express our appreciation to the staff of National Astronomical Observatory of Japan for
their support during our IRCS observation, and also to the staff of IRTF during our
NSFCam observation. M.T. and M.G. acknowledge financial support from the Japan Society
for the Promotion of Science. This publication makes use of data products from the Two
Micron All Sky Survey, which is a joint project of the University of Massachusetts and
the Infrared Processing and Analysis Center/California Institute of Technology, funded
by the National Aeronautics and Space Administration and the National Science Foundation.

\clearpage

\clearpage


\clearpage

\begin{deluxetable}{cccrrrrcc}
 \tabletypesize{\scriptsize}
 \tablecaption{The Source List\label{tb:t1}}
 \tablecolumns{9}
 \tablewidth{0pt}
 \tablehead{
 \colhead{Name} &
 \colhead{R.A.\tablenotemark{a}} &
 \colhead{DEC.\tablenotemark{a}} &
 \colhead{\ji\tablenotemark{b}} &
 \colhead{\hi\tablenotemark{b}} &
 \colhead{\ki\tablenotemark{b}} &
 \colhead{\li\tablenotemark{b}} &
 \colhead{2MASS} &
 \colhead{X-ray\tablenotemark{c}} \\
 \colhead{} &
 \colhead{} &
 \colhead{} &
 \colhead{mag} &
 \colhead{mag} &
 \colhead{mag} &
 \colhead{mag} &
 \colhead{Identification} &
 \colhead{Identification}
 }
 \startdata
 IRS1 & 05:35:17.415 & $-$04:59:57.24 & 17.8 & 14.7 & 13.0 & \nodata & 0535174-045957 & X-6\phn \\
 IRS2 & 05:35:16.168 & $-$05:00:02.58 & 17.0 & 14.1 & 12.3 & \nodata & 0535161-050002 & \nodata \\
 IRS3\tablenotemark{d} & 05:35:18.275 & $-$05:00:33.93 & $>$19.6 & 18.0 & 13.2 & 9.17 & 0535183-050033\tablenotemark{e} & \nodata \\
 IRS4 & 05:35:17.736 & $-$05:00:31.07 & 15.4 & 13.2 & 11.7 & 10.5 & 0535177-050031 & X-7\phn \\
 IRS5\tablenotemark{d} & 05:35:18.340 & $-$05:00:33.01 & $>$19.6 & 15.8 & 11.4 & 7.86 & 0535183-050033\tablenotemark{e} & X-8c \\
 IRS6 & 05:35:15.265 & $-$05:00:33.47 & 13.3 & 12.7 & 12.5 & \nodata & 0535152-050033 & X-2\phn \\
 IRS7 & 05:35:15.837 & $-$05:00:36.34 & 12.3 & 11.8 & 11.7 & \nodata & 0535158-050036 & \nodata \\
 IRS8 & 05:35:19.980 & $-$05:01:02.64 & $>$19.6 & $>$18.8 & 14.4 & \nodata & 0535199-050102 & X-12\\
 IRS9 & 05:35:15.463 & $-$05:01:12.59 & 14.2 & 13.6 & 13.4 & \nodata & 0535154-050112 & X-3\phn \\
 \enddata
 \tablenotetext{a}{The positions are determined from the IRCS \ki-band image in the equinox J2000.0.}
 \tablenotetext{b}{All magnitudes are in the CIT system.}
 \tablenotetext{c}{The X-ray counterpart with the source number in \citet{tsuboi01}.}
 \tablenotetext{d}{Associated with MMS2.}
 \tablenotetext{e}{IRS3 and IRS5 are not resolved in the 2MASS data.}
\end{deluxetable}


\begin{thebibliography}{}
 \bibitem[Aso et al.(2000)]{aso00} Aso, Y., Tatematsu, K., Sekimoto, Y., Nakano, T.,
			   Umemoto, T., Koyama, K., \& Yamamoto, S. 2000, \apjs, 131,
			   465
 \bibitem[Aspin, Sandell, \& Russell(1994)]{aspin94} Aspin, C., Sandell, G., \& Russell,
			   A. P. G. 1994, \aaps, 106, 165
 \bibitem[Bally(1993)]{bally93} Bally, J., Devine, D., Herald, M., \& Rauscher,
			   B. J. 1993, \apj, 418, 75
 \bibitem[Barsony(1994)]{barsony94} Barsony, M. 1994, in ASP Conf. Ser. 65, Clouds,
			   Cores, and Low Mass Stars, ed. D. P. Clemens and R. Barvainis
			   (San Francisco: ASP), 197
 \bibitem[Bertin \& Arnouts(1996)]{bertin96} Bertin, E. \& Arnouts, S. 1996, \aap, 113, 393
 \bibitem[Carpenter(2001)]{carpenter01} Carpenter, J. M. 2001, \aj, 121, 2851
 \bibitem[Chini et al.(1997)]{chini97} Chini, R., Reipurth, B., Ward-Thompson, D.,
			   Bally, J., Nyman, L-\AA, Sievers, A., \& Billawala, Y. 1997,
			   \apj, 474, L135
 \bibitem[Davis et al.(1994)]{davis94} Davis, C. J., Dent, W. R. F., Matthews, H. E.,
			   Aspin, C., and Lightfoot, J. F. 1994, \mnras, 266, 933
 \bibitem[Eisl\"{o}ffel(2000)]{eisloeffel00} Eisl\"{o}ffel, J. 2000, \aap, 354, 236
 \bibitem[Hodapp \& Ladd(1995)]{hodapp95} Hodapp, K. W. \& Ladd, E. F. 1995, \apj, 453,
			   715
 \bibitem[Hodapp(1998)]{hodapp98} Hodapp, K-W. 1998, \apj, 500, L183
 \bibitem[Johnstone \& Bally(1999)]{johnstone99} Johnstone, D. \& Bally, J. 1999, apj, 510, L49
 \bibitem[Kobayashi et al.(2000)]{kobayashi00} Kobayashi, N., et al.
			   2000, Proc. of SPIE, 4008, 1056
 \bibitem[Lada \& Adams(1992)]{lada92} Lada, C. J., \& Adams, F. C. 1992, \apj, 393, 278
 \bibitem[Leggett \& Denault(1996)]{leggett96} Leggett, S. \& Denault, T. 1996,
			   NSFCAM 256$\times$256 InSb Infrared Array Camera User's
			   Guide, Version 3, NASA Infrared Telescope Facility
 \bibitem[Lis et al.(1998)]{lis98} Lis, D. C., Serabyn, E., Dowell, C. D., Benford,
			   D. J., Phillips, T. G., Hunter, T. R., \& Wang, N. 1998,
			   \apj, 509, 299
 \bibitem[Meyer et al(1997)]{meyer97} Meyer, M. R., Calvet, N., \& Hillenbrandt,
			   L. A. 1997, \aj, 114, 288
 \bibitem[Reipurth, Bally, \& Devine(1997)]{reipurth97} Reipurth, B., Bally, J., \&
			   Devine, D. 1997, \aj, 114, 2708
 \bibitem[Reipurth, Rodr\'{\i}guez \& Chini(1999)]{reipurth99} Reipurth, B.,
			   Rodr\'{\i}guez, L. F., \& Chini, R. 1999, \aj, 118, 983
 \bibitem[Tokunaga(1998)]{tokunaga98} Tokunaga, A. T., et al. 1998, Proc. of SPIE, 3354,
			   512
 \bibitem[Tokunaga(2000)]{tokunaga00} Tokunaga, A. T. 2000, in Allen's Astrophysical
			   Quantities, ed. A. N. Cox, (4th ed.; New York:
			   Springer-Verlag), 143
 \bibitem[Tsuboi et al.(2001)]{tsuboi01} Tsuboi, Y., Koyama, K., Hamaguchi, K.,
			   Tatematsu, K., Sekimoto, Y., Bally, J., \& Reipurth, B.
			   2001, \apj, 554, 734
 \bibitem[Tsujimoto et al.(2001)]{tsujimoto01} Tsujimoto, M., Koyama, K., Tsuboi, Y.,
			   Goto, M., and Kobayashi, N. 2001, \apj, 556, in press
 \bibitem[Yu, Bally, \& Devine(1997)]{yu97} Yu, K. C., Bally, J., \& Devine, D. 1997, \apj,
                           485, L45
 \bibitem[Yun et al.(1997)]{yun97} Yun, J. L., Clemens, Dan P., Morelra, M. C., and
			   Santos, N. C. 1997, \apj, 479, L71
\end{thebibliography}
\end{document}